\documentclass{article}

\usepackage{arxiv}

\usepackage[utf8]{inputenc} 
\usepackage[T1]{fontenc}    
\usepackage{hyperref}       
\usepackage{url}            
\usepackage{booktabs}       
\usepackage{amsfonts}       
\usepackage{nicefrac}       
\usepackage{microtype}      
\usepackage{lipsum}
\usepackage{graphicx}

\usepackage{amsmath}
\usepackage{mathrsfs}

\usepackage{physics}
\usepackage{hyperref}
\usepackage{bm}

\usepackage{xcolor}
\usepackage{soul}

\usepackage{booktabs}
\usepackage{multirow}
\usepackage{natbib}
\usepackage{physics}
\renewcommand{\vec}{\bm}
\newcommand{\mat}{\vec}
\DeclareMathOperator{\E}{E}
\DeclareMathOperator{\Var}{Var}
\DeclareMathOperator{\AVar}{AVar}
\DeclareMathOperator{\Cov}{Cov}
\DeclareMathOperator{\Cor}{Cor}
\DeclareMathOperator{\BIC}{BIC}

\DeclareMathOperator{\vech}{vech}
\newcommand{\loss}{\mathcal{L}}
\newcommand{\risk}{\mathcal{R}}
\renewcommand{\P}{\mathbb{P}}
\newcommand{\pto}{\overset{\mathrm{p}}\to}

\newcommand{\dto}{\overset{d}\to}

\newtheorem{theorem}{Theorem}

\usepackage{siunitx}

\DeclareMathOperator*{\argmin}{arg\,min}

\title{Jointly modeling multiple endpoints for efficient treatment effect estimation in randomized controlled trials}

\author{
 Jack M.~Wolf \\
  Division of Biostatistics \& Health Data Science \\
  University of Minnesota \\
  Minneapolis, Minnesota, U.S.A.\\
  \texttt{wolfx681@umn.edu} \\
   \And
 Joseph S. Koopmeiners \\
  Division of Biostatistics \& Health Data Science \\
  University of Minnesota \\
  Minneapolis, Minnesota, U.S.A.\\
  \And
 David M. Vock \\
  Division of Biostatistics \& Health Data Science \\
  University of Minnesota \\
  Minneapolis, Minnesota, U.S.A.\\
}

\begin{document}
\maketitle
\begin{abstract}
Randomized controlled trials are the gold standard for evaluating the efficacy of an intervention. However, there is often a trade-off between selecting the most scientifically relevant primary endpoint versus a less relevant, but more powerful, endpoint. For example, in the context of tobacco regulatory science many trials evaluate cigarettes per day as the primary endpoint instead of abstinence from smoking due to limited power. Additionally, it is often of interest to consider subgroup analyses to answer additional questions; such analyses are rarely adequately powered.  In practice, trials often collect multiple endpoints. Heuristically, if multiple endpoints demonstrate a similar treatment effect we would be more confident in the results of this trial. However, there is limited research on leveraging information from secondary endpoints besides using composite endpoints which can be difficult to interpret. In this paper, we develop an estimator for the treatment effect on the primary endpoint based on a joint model for primary and secondary efficacy endpoints. This estimator gains efficiency over the standard treatment effect estimator when the model is correctly specified but is robust to model misspecification via model averaging. We illustrate our approach by estimating the effect of very low nicotine content cigarettes on the proportion of Black people who smoke who achieve abstinence and find our approach reduces the standard error by 27\%.
\end{abstract}


\section{Introduction}

Randomized controlled trials (RCTs) are the gold standard for evaluating the efficacy of an intervention; however, they can be costly, time-consuming, and face challenges enrolling a sufficient sample size.
Moreover, there is often a trade-off between selecting the most relevant primary endpoint versus a more powerful, but less relevant endpoint.
Additionally, secondary analyses of RCTs often target subgroup treatment effects in priority populations; such analyses often have small sample sizes and limited power.
Thus, it is of interest to develop more efficient methods RCTs that are likely to succeed with fewer participants.

We seek to gain precision by leveraging information found in secondary endpoints. 
Secondary endpoints are regularly measured in RCTs with the goal of understanding the effect of a treatment on complex phenomenon such or disease which cannot be fully reduced to one, single primary endpoint. 
Intuitively, investigators may be more confident in their conclusions if they observe a strong signal on the primary and secondary endpoints versus if they had only observed a signal on the primary endpoint.
In this manuscript, we formalize this heuristic thinking and develop a statistical estimator that integrates the information from secondary endpoints to better estimate the effect on the primary endpoint.

Our work is in part motivated by recent studies of very low nicotine content (VLNC) cigarettes in tobacco regulatory science. 
RCTs have consistently shown that people who smoke (PWS) who are randomized to receive VLNC cigarettes smoke significantly fewer cigarettes per day (CPD), and have lower biomarkers of nicotine and toxicant exposure, and lower measures of dependence than those randomized to receive normal nicotine content (NNC) cigarettes \citep{donny_randomized_2015, hatsukami_effect_2018, white_preliminary_2022, hatsukami_reduced_2024}.
However, these trials have not been powered to detect an effect on abstinence from smoking, a rare binary endpoint, or to estimate effects within priority populations. 
For example, Black PWS have been identified as a priority population due to being disproportionately burdened by the health effects of smoking and it is crucial to understand the effect of VLNC cigarettes in this population. 
We wish to leverage available information about CPD, biomarkers of nicotine and toxicant exposure, and dependence to obtain a precise estimate of the effect on abstinence.

Recently, \cite{chen_improving_2022, chen_efficient_2023} leveraged information from secondary endpoints to achieve more precise estimation of covariates' associations with a primary endpoint.
However, their work was motivated by observational data, and we have found that their class of estimators cannot gain efficiency on treatment effect estimators within RCTs due to the independence between treatment assignment and baseline covariates \citep{wolf_commentary_2024}.
Instead, we develop a novel estimator that draws from the structural equation modeling literature to leverage information from multiple endpoints. 
Structural equation models (SEMs) are extensions of factor analytic models that seek to understand the correlation structure of several endpoints by relating them to latent constructs and to estimate structural relationships between these constructs and other covariates \citep{joreskog_general_1970, joreskog_estimation_1975, beran_structural_2010}. 
Distinctly, our interest lies in leveraging the underlying latent structure within the data to achieve a more precise estimator for the primary outcome, rather than understanding latent relationships, which is the traditional goal of many SEMs and requires strong, unverifiable assumptions about the latent structure of the data.
The SEM approach results in more precise estimation but requires a correctly specified model. To improve robustness, we propose a model averaging-based estimator that leverages the SEM to gain efficiency when the underlying relationship can be captured appropriately from an SEM but is robust to model misspecification. 

The rest of the paper proceeds as follows. In Section \ref{sec:jm} we consider approaches to jointly model primary and secondary endpoints and introduce our proposed estimators.
Section \ref{sec:sims} presents simulation results comparing our estimators to standard estimators that do not leverage secondary endpoints and Section \ref{sec:application} applies our estimators to a recent trial of VLNC cigarettes. 
Finally, we conclude with a discussion in Section \ref{sec:discussion}.

\section{Approaches to Joint Modeling}\label{sec:jm}
\begin{figure}
    \centering
    \includegraphics[width=\linewidth]{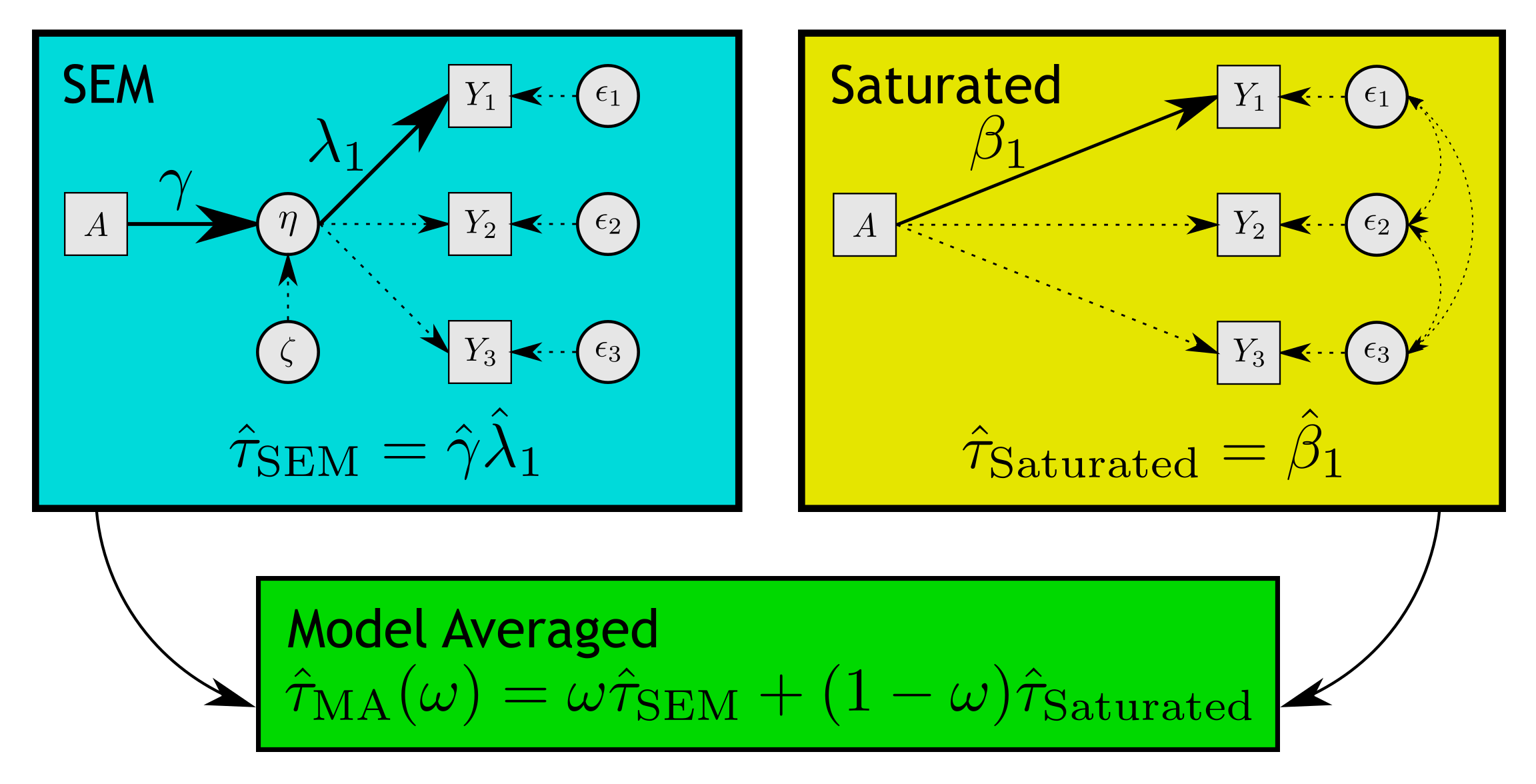}
    \caption{Conceptual estimation approach with three endpoints. Observed variables are represented in squares while latent variables are in circles. The structural equation model (SEM) assumes there is no residual correlation between endpoints after accounting for the the latent variable $\eta$ while the saturated model does not restrict the endpoint correlation structure. These models and their corresponding point estimates are then averaged across for a more robust point estimate.}
    \label{fig:model-concept}
\end{figure}

We incorporate information from secondary endpoints by proposing a joint model for all endpoints. 
These parametric models identify the average treatment effect (or other estimands of interest) and suggest consistent estimators based on maximum likelihood theory. 
A conceptual diagram of this approach is provided in Figure \ref{fig:model-concept}. We will consider two joint models for the endpoints: one saturated model and one SEM. Using model averaging, we combine the resultant point estimates from each model to obtain an estimator that is robust to SEM misspecification.

\subsection{Notation and Preliminaries}\label{subsec:notation}

We assume that observations are of the form $\{A_i,\vec{Y}_i\}_{i=1}^n$ and independent and identically distributed where $A_i$ is a binary treatment indicator and $\vec{Y}_i = (Y_{i,1},\ldots,Y_{i,P})^T$ where $Y_{i,p}$ is a measurement of the $p$th endpoint for $P\ge3$ total endpoints (noting that the model introduced in Section \ref{subsec:sem} is either under-identified or saturated with only $1$ or $2$ endpoints, respectively). 
Endpoints may be numerical, binary, or ordinal.
Without loss of generality, we let $Y_{i,1}$ be the primary endpoint.
The estimand of interest is the average treatment effect on the primary endpoint: $\tau_1 = \E(Y_{i,1}|A_i=1)-\E(Y_{i,1}|A_i=0)$.

\subsection{Saturated Mean Model}\label{subsec:saturated}

We first consider a saturated joint model for all endpoints. 
In particular, for numerical endpoints we posit a joint Gaussian model (top left panel of Figure \ref{fig:model-concept}):
\begin{equation}
    \vec{Y}_i|A_i = N\left(\vec\alpha + \vec\beta A_i, \vec\Sigma \right)
\end{equation}
where $\vec\alpha$ and $\vec\beta$ are $p\times1$ vectors of intercepts and treatment effects, respectively, for each endpoint and $\vec\Sigma$ is a $p\times p$ endpoint covariance matrix.
The implied treatment effect estimator is $\hat\tau_\text{Saturated}=\hat\beta_1$ where $\hat\beta_1$ is the maximum likelihood estimate (MLE) for $\beta_1$. 
It can be shown that the resultant treatment effect estimator is the difference in sample means and does not include any information from secondary endpoints:
$$\hat\tau_\text{Saturated} = \frac{\sum_{i=1}^n A_i Y_{i,1}}{\sum_{i=1}^nA_i} - \frac{\sum_{i=1}^n (1-A_i) Y_{i,1}}{\sum_{i=1}^n(1-A_i)}.$$
Similar results occur with categorical endpoints.
Instead, to gain efficiency from secondary endpoints, we must impose constraints on the model.

\subsection{Structural Equation Model}\label{subsec:sem}

When there are three or more endpoints, we can impose mean--covariance constraints inspired by structural equation modeling to gain efficiency in treatment effect estimation.
Specifically, we will assume that the endpoints are affected by the treatment through one latent variable, $\eta_i$.

For multivariate Gaussian endpoints, we consider the following SEM with one latent variable (top right panel of Figure \ref{fig:model-concept}):
\begin{align}
    \label{eqn:sem}
    \begin{split}
    \vec{Y}_i|\eta_i &\sim N\left\{\vec\nu + \vec\lambda \eta_i, \operatorname{diag}(\theta_1, \ldots, \theta_P) \right\} \\
    \eta_i|A_i &\sim N(\gamma A_i, 1),
    \end{split}
\end{align}
where $\vec\nu$ and $\vec\lambda$ are both $P\times 1$ vectors. 
This imposes the following model for the observed data:
\begin{equation}
    \label{eqn:sem-observed}
    \vec{Y}_i|A_i \sim N\left\{\vec\nu + \gamma  \vec\lambda A_i,\operatorname{diag}(\theta_1, \ldots, \theta_P) + \vec\lambda\vec\lambda^T   \right\}.
\end{equation}
Importantly, under this likelihood we have
\begin{align*}
     \E(Y_{i,p}|A_i=1)-\E(Y_{i,p}|A_i=0) = \gamma\lambda_p\quad \text{for all}\quad p. 
\end{align*}
It follows that, for $\lambda_p\neq0$, 
$$\gamma = \left\{\E(Y_{i,p}|A_i=1)-\E(Y_{i,p}|A_i=0)\right\}/\lambda_p$$
and therefore
\begin{align*}
    \E(Y_{i,1}|A_i=1)-\E(Y_{i,1}|A_i=0) =  \lambda_1\left\{\E(Y_{i,p}|A_i=1)-\E(Y_{i,p}|A_i=0)\right\}/\lambda_p.
\end{align*}
That is, the average treatment effect on the primary endpoint is a function of the treatment effect on any given secondary endpoint, provided that the latent variable affects the secondary endpoint.
Consequently, the treatment effect estimator given by the MLE from this model, $\hat\tau_\text{SEM} = \hat\gamma\hat\lambda_1$, may gain efficiency over the $\hat\tau_\text{Saturated}$ by borrowing strength from treatment effects for the secondary endpoints and correlations between the primary and secondary endpoints.
We note that, by the delta method, the asymptotic variance $\hat\tau_\text{SEM}$ is
\begin{equation}
    \AVar(\hat\tau_\text{SEM}) = (\lambda_1, \gamma)\Cov(\hat\gamma, \hat\lambda_1)(\lambda_1, \gamma)^T,
\end{equation}
which can be estimated using a plug in estimator.

This model can then be extended to accommodate binary and ordinal endpoints by adding additional latent variables to represent continuous realizations of these endpoints (i.e., probit regression).
Specifically, we let $Y_{i,j}$ be the observed categorical realization of the latent Gaussian endpoint, $Y_{i,j}^*$, where
$$
    Y_{i,j} = \begin{cases}0 & Y_{i,j}^* \le 0 \\
 1 & 0 < Y_{i,j}^* \le a_1 \\ 
    \vdots \\ 
    k & a_k < Y_{i,j}^*\end{cases}.
$$
Details of the observed data likelihood calculation and maximization are provided in Appendix \ref{sec:sem-likelihood}.

We note that for a binary primary endpoint, the marginal observed likelihood is
$$Y_{i,1}|A_i \sim \operatorname{Bernoulli}\left\{p_i = \Phi\left(\frac{\nu_1 + \gamma\lambda_1A_i}{\sqrt{1+\lambda_1^2}}\right) \right\}, $$
and the average treatment effect is
$$\E(Y_{i,1}|A_i=1)-\E(Y_{i,1}|A_i=0) = \Phi\left(\frac{\nu_1 + \gamma\lambda_1}{\sqrt{1+\lambda_1^2}} \right) - \Phi\left(\frac{\nu_1}{\sqrt{1+\lambda_1^2}} \right), $$
where $\Phi$ is the Gaussian cumulative distribution function.
By the delta method, the asymptotic variance of the treatment effect estimator is
$$
\AVar\left\{\Phi\left(\frac{\hat\nu_1 + \hat\gamma\hat\lambda_1}{\sqrt{1+\hat\lambda_1^2}} \right) - \Phi\left(\frac{\hat\nu_1}{\sqrt{1+\hat\lambda_1^2}} \right) \right\} = j(\gamma, \nu_1, \lambda_1) \Cov(\hat\gamma, \hat\nu_1, \hat\lambda_1)j(\gamma, \nu_1, \lambda_1)^T,$$ 
where
\begin{align*}
    j(\gamma, \nu_1, \lambda_1)^T &= 
    \phi\left(\frac{\nu_1 + \gamma\lambda_1}{\sqrt{1+\lambda_1^2}}\right)
    \begin{pmatrix}
        {\lambda_1}/{\sqrt{1+\lambda_1^2}} \\
        {1}/{\sqrt{1+\lambda_1^2}} \\
        \dfrac{\gamma-\nu_1\lambda_1}{(1+\lambda_1^2)^{3/2}}
    \end{pmatrix}
    -
    \phi\left(\frac{\nu_1}{\sqrt{1+\lambda_1^2}}\right)
    \begin{pmatrix}
        0 \\
        {1}/{\sqrt{1+\lambda_1^2}} \\
        \dfrac{-\nu_1\lambda_1}{(1+\lambda_1^2)^{3/2}}
    \end{pmatrix},
\end{align*}
and $\phi$ is the standard Gaussian probability density function.
If the primary endpoint is ordinal with three or more levels, other estimands must be considered.
Possible estimands include the probit regression coefficient, 
$$\tau_{1,\text{probit}} = \frac{\E(Y_{i,1}^*|A_i=1)-\E(Y_{i,1}^*|A_i=0)}{\sqrt{\Var(Y_{i,1}^*|A_i)}},$$ 
and the concordance probability,
$$\tau_{1,\text{concordance}} = \Pr(Y_{i,1}>Y_{j,1}|A_i=1,A_j=0) + \frac12\Pr(Y_{i,1}=Y_{j,1}|A_i=1,A_j=0).$$

\subsection{Model Averaging Based Estimators}\label{subsec:model-averaging}
We have considered two treatment effect estimators: the standard difference in sample means estimator which is the MLE under the saturated model: $\hat\tau_\text{Saturated}$, and the MLE under the SEM: $\hat\tau_{\text{SEM}}$.
To protect against biases that may stem from model misspecification under the SEM while still gaining efficiency when the SEM is correctly specified, we consider estimators that use model averaging:
\begin{equation}
 \hat\tau_\text{MA}(\omega) = \omega\hat\tau_\text{SEM}+(1-\omega)\hat\tau_\text{Saturated}   
\end{equation}
for $\omega\in[0,1]$, where $\omega$ is a data-driven weight estimated from the data.
We  consider two methods of estimating optimal weights. 

First, we use weights based on the Bayesian information criterion (BIC) with
\begin{equation}
\hat\omega_\text{BIC} = \left[\exp\left\{\frac12\left(\BIC_\text{SEM}-\BIC_\text{Saturated} \right) \right\} +1\right]^{-1}
\end{equation}
and $\hat\tau_\text{BIC}=\hat\tau_\text{MA}(\hat\omega_\text{BIC})$. 
Here $\BIC_\text{SEM}$ and $\BIC_\text{Saturated}$ are the BICs for full models that include all $P$ endpoints.
This weight appeals to the consistent model selection properties of the BIC and, under a Bayesian lens with noninformative priors, can be interpreted as an approximation of the posterior probability that the SEM is correct \citep{hjort_frequentist_2003}. 

Second, we facilitate model averaging through ensemble Super Learning \citep{van_der_laan_super_2007}.
Briefly, we estimate the cross-validated mean squared prediction error for the primary endpoint (ignoring the secondary endpoints) for both models. 
Then, we identify the weights $\hat\omega_\text{SL}$ and $1 - \hat\omega_\text{SL}$ that minimize the cross validated mean squared error (MSE) across all weighted combinations of the two models:
\begin{equation}
    \hat\omega_\text{SL} = \argmin_{\omega\in[0,1]} \sum_{i=1}^n \left[Y_{i,1} - \{\omega\hat{Y}_{i,\text{SEM}} + (1 - \omega)\hat{Y}_{i,\text{Saturated}}\}\right]^2
\end{equation}
where $\hat{Y}_{i,\text{SEM}}$ and $\hat{Y}_{i,\text{Saturated}}$ are estimates of $Y_{i,1}$ under a SEM and saturated model fit to a dataset not including subject $i$.
Additional details are provided in Appendix \ref{sec:super}.

We approximate the sampling distribution of both model averaging estimators using the nonparametric bootstrap. 
Approximate $1-\alpha$ confidence interval (CI) bounds are obtained via the $\alpha/2$ and $(2-\alpha)/2$ percentiles of the bootstrapped sampling distribution, and Wald test statistics are obtained by dividing the point estimate by the bootstrapped standard error.

\subsection{Theoretical Large Sample Results}

Herein we present several asymptotic results considering scenarios in which all endpoints are multivariate Gaussian, noting that binary and ordinal probit models for non-Gaussian data can be accommodated with additional notation and identifiability constraints. 
Formal proofs are provided in Appendices \ref{sec:pf-sem-efficiency} and \ref{sec:pf-ma-consistent}.

Our first result states that the SEM estimator is more efficient than the saturated estimator when the SEM is correctly specified. 
Heuristically, this occurs because there is additional information about $\tau$ encoded in the covariance between endpoints; ignoring this structure results in a less efficient estimator and an inefficient use of all available information. 

\begin{theorem}[Efficiency of SEM Estimator]\label{th:efficiency}
    If the SEM is correctly specified such that
    $$  \vec{Y}_i|A_i \sim N\{\vec\nu + \vec\lambda\gamma A_i,\operatorname{diag}(\theta_1, \ldots, \theta_P) + \vec\lambda\vec\lambda^T\},$$
    then
    $\AVar(\hat\tau_\text{SEM}) \le \AVar(\hat\tau_\text{Saturated})$,
    where $\AVar(X)$ denotes the asymptotic variance of $X$.
\end{theorem}

The proof proceeds by recognizing $\hat\tau_\text{SEM}$ as a function of the solution to a set of quadratic estimating equations versus $\hat\tau_\text{Saturated}$, which is a function of the solution to linear estimating equations \citep{carroll_comparison_1982}. Asymptotic variance formulae for both estimators are then obtained using $M$-estimation theory.
This framework also allows for exploration of the approximate bias and variance of $\hat\tau_\text{SEM}$ when the SEM is misspecified.

Next, we show that when model averaging, the totality of the weight is asymptotically placed on the ``correct'' model, leading to consistency of both model averaging estimators as well as variance reduction when the SEM is correct.

\begin{theorem}[Consistency of Model Averaging Estimators]
    Both $\hat\tau_\text{BIC}$ and $\hat\tau_\text{SL}$ are consistent estimators for $\tau_1$.
    Moreover, $\AVar(\hat\tau_\text{BIC}) = \AVar(\hat\tau_\text{SL}) = \AVar(\hat\tau_\text{SEM})$ when the SEM is correctly specified.
\end{theorem}

This result is proven for $\hat\tau_\text{BIC}$ by realizing that $\hat\omega_\text{BIC}$ is approximately a function of a (noncentral) $\chi^2$ random variable minus a $\log(n)$ term. 
If the SEM is correctly specified, the noncentrality parameter is equal to $0$ and the $\log(n)$ penalty leads $\omega_\text{BIC}$ to converge in probability to $0$.
However, when the SEM is misspecified, the noncentrality parameter grows linearly in $n$ and dominates the $\log(n)$ penalty parameter causing $\hat\omega_\text{BIC}$ to converge to $1$. 
Results for $\hat\tau_\text{SL}$ are a direct result of the Super Learner having asymptotically equivalent performance to oracle estimators which select the optimal weighted combination of estimators to minimize the expected value of the loss function \citep{van_der_laan_super_2007}. 
Because the estimated mean model under the SEM is unbiased and efficient when the SEM is correct, the oracle estimator will place all weight on the $\hat\tau_\text{SEM}$, minimizing the MSE.
Conversely, when the SEM is incorrect, the saturated model is the only unbiased estimator for the mean and must receive weight $1$ to minimize the MSE in large samples.

\section{Simulation Studies}\label{sec:sims}

We evaluate the small sample properties of our proposed estimators over a variety of scenarios. We vary the endpoint types and correlation structure while holding the (standardized) treatment effects and sample size fixed. 
Across all simulations, we summarize the bias, standard error, root mean squared error (RMSE), coverage, and power of each estimator over \num[group-separator={,}]{1000} independent Monte-Carlo simulations.

\subsection{Simulation A: Three Gaussian Endpoints, SEM Correctly Specified}

We consider three multivariate Gaussian endpoints under a global null hypothesis of no treatment effect on any endpoint and under the alternative with respective (standardized) average treatment effects of $0.25$, $0.35$, and $0.3$. 
Under both the null and alternative  mean structures, we manipulate the endpoint correlation matrix across a range of correlations consistent with the specified SEM.

Under the alternative, we consider a grid of $\Cov(Y_{i,1},Y_{i,2}|A_i)\in\{0.20, 0.25, \ldots, 0.70\}$ with the mean structure specified above and $\Var(Y_{i,j}|A_i)=1$ for $j=1,2,3$ by setting $\gamma = \sqrt{0.25\times0.35/\Cov(Y_{i,1},Y_{i,2}|A_i)}$, which in turn manipulates both $\Cov(Y_{i,1},Y_{i,3}|A_i)$ and $\Cov(Y_{i,2},Y_{i,3}|A_i)$ as well.
We evaluate performance using the same correlation matrices under the global null hypotheses, noting that these models correspond to correctly specified SEMs with $\gamma=0$.

We fix the total sample size at $n=250$ with $125$ subjects in each treatment arm. This gives $80\%$ power to detect an effect on $Y_{i,2}$ under the alternative but only $50\%$ power to detect an effect on $Y_{i,1}$.

\begin{figure}[htp]
    \centering
    \includegraphics[width=\linewidth]{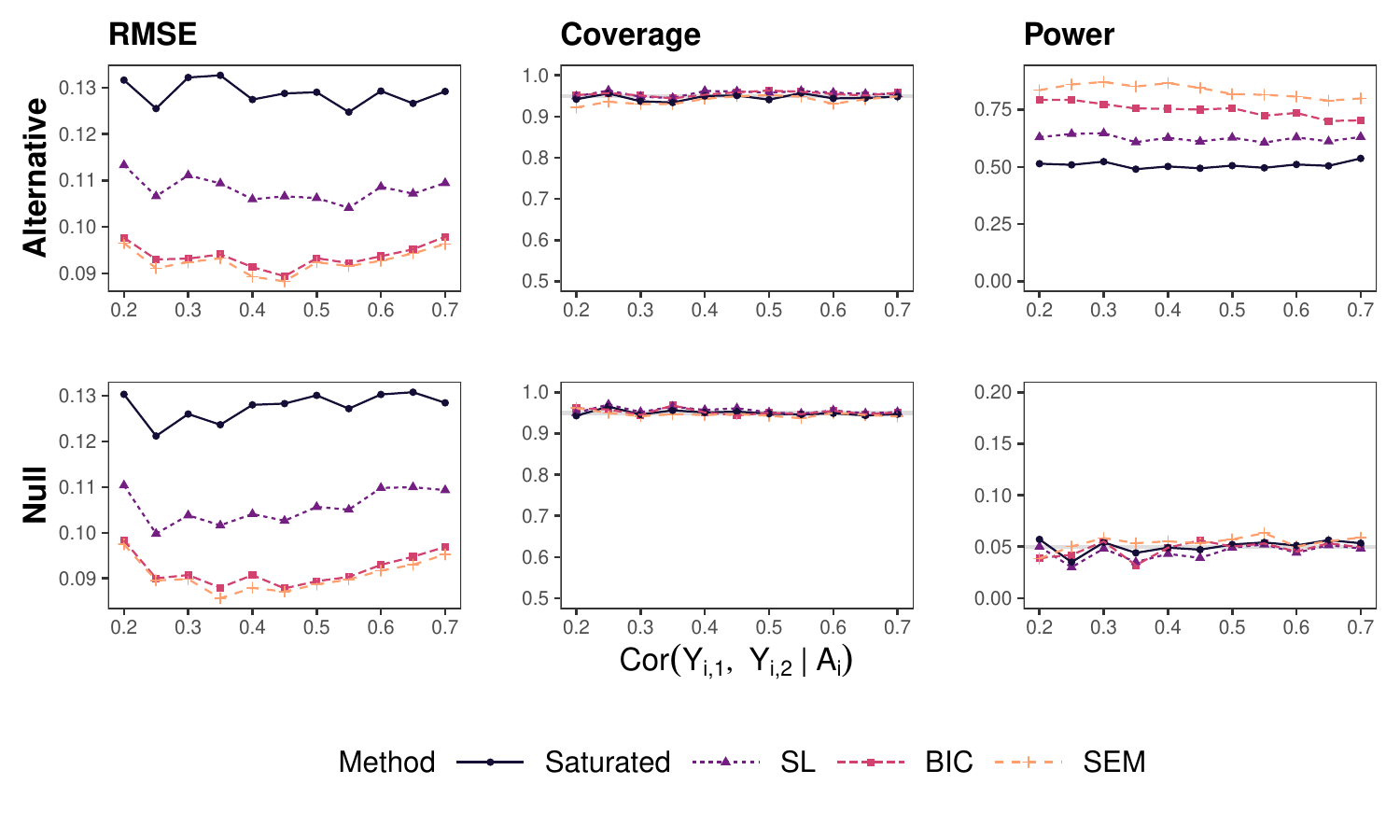}
    \caption{
        Simulation A: Root mean squared error (RMSE), coverage, and power under correctly specified structural equation model. 
    }
    \label{fig:sim-a}
\end{figure}

When the SEM is correct (Figure \ref{fig:sim-a}), all estimators are approximately unbiased.
The estimators that use data from secondary endpoints (SEM and both model-averaging estimators) gain efficiency over the saturated estimator resulting in lower RMSEs across all explored correlations. 
Of note, there is little association between correlation and efficiency gain.
Under the alternative hypothesis, these efficiency gains translate into increases in power with respective empirical rejection rates of approximately $60\%$, $75\%$, and $85\%$ for  $\hat\tau_\text{SL}$, $\hat\tau_\text{BIC}$, and $\hat\tau_\text{SEM}$ when $\Cor(Y_{i,1},Y_{i,2}|A_i)=0.35$.
All estimators maintain 95\% coverage and control the Type~I error under the null hypothesis.  
$\hat\tau_\text{BIC}$ performs similarly to $\hat\tau_\text{SEM}$, placing nearly all weight on the SEM in most simulations. 
In contrast, $\hat\tau_\text{SL}$ places less weight on the SEM estimator and has smaller efficiency gains versus the saturated model.

\subsection{Simulation B: Three Gaussian Endpoints, SEM Misspecified}

To evaluate the performance of the SEM estimator under model misspecification, we explore three scenarios where the SEM is misspecified.
Under the first, we assume the same mean structure as under the alternative hypothesis in the previous simulation. 
Then, in the second scenario, we evaluate performance with a null treatment effect for the primary but not the secondary endpoints with respective average treatment effects of $0, 0.35$, and $0.3$.
Finally, in the third scenario, we consider a global null hypothesis.
In the first two cases, we perturb the endpoint correlation in a way that was incompatible with a SEM. 
Specifically, we set
$$\Cov(Y_{i,1}, Y_{i,2}, Y_{i,3}|A_i) = \begin{bmatrix} 1 & 0.35s & 0.30s \\  & 1 & 0.42 \\  & & 1 \end{bmatrix}$$ 
and vary $s \in \{0, 0.25, \ldots, 2\}$. 
With the given mean structure, this represents a SEM with $\gamma = 0.5$ if and only if $s = 1$ under the alternative scenario ($\vec\lambda=(0.5, 0.7, 0.6)^T$) and if and only if $s = 0$ under the null scenario ($\vec\lambda=(0, 0.7, 0.6)^T$).
In the third scenario, we set 
$$\Cov(Y_{i,1}, Y_{i,2}, Y_{i,3}|A_i) = \begin{bmatrix} 1 & 0.35s & 0.30 \\  & 1 & 0.42 \\  & & 1 \end{bmatrix}.$$
Again, we vary $s \in \{0, 0.25, \ldots, 2\}$; the SEM is correctly specified if and only if $s = 1$ ($\gamma = 0$ and $\vec\lambda = (0.5, 0.7, 0.6)^T)$.

\begin{figure}[htbp]
    \centering
    \includegraphics[width=\linewidth]{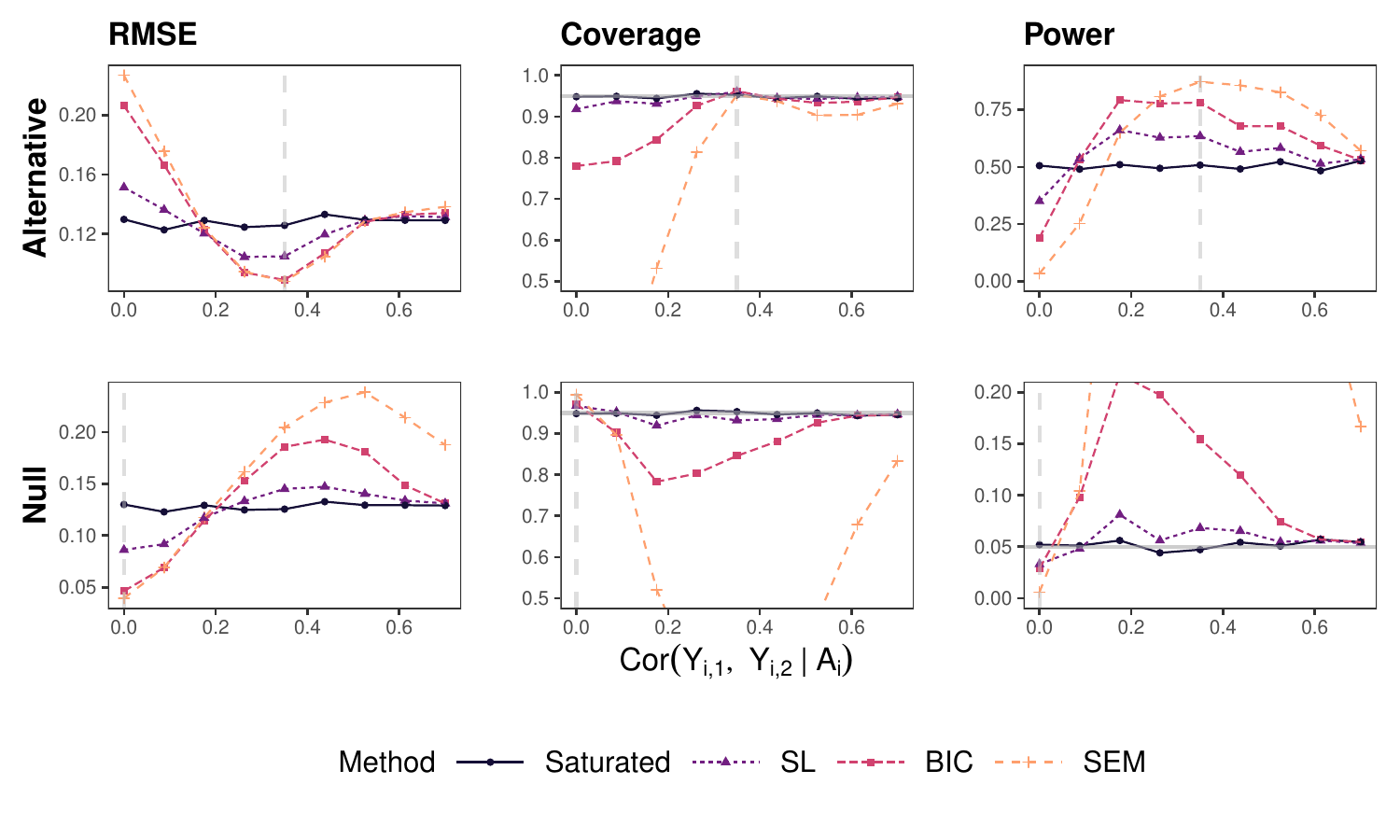}
    \caption{
        Simulation B1: Root mean squared error (RMSE), coverage, and power when the structural equation model is misspecified at all but one endpoint correlation. 
        A dashed vertical line indicates the correlation at which the structural equation is correctly specified.
    }
    \label{fig:sim-b}
\end{figure}

\begin{figure}[htbp]
    \centering
    \includegraphics[width=\linewidth]{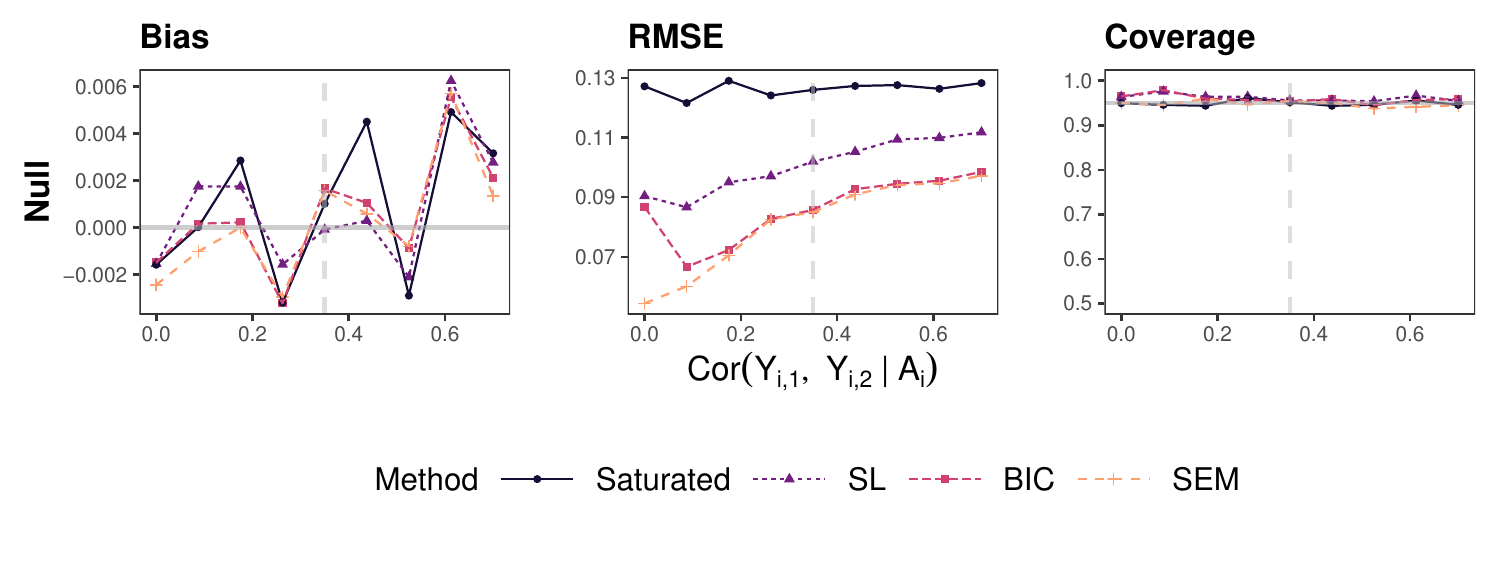}
    \caption{
        Simulation B2: Bias, root mean squared error (RMSE), and coverage when the structural equation model is misspecified at all but one endpoint correlation under the global null hypothesis. 
        A dashed vertical line indicates the correlation at which the structural equation is correctly specified.
    }
    \label{fig:sim-b2}
\end{figure}

In the first two scenarios of Simulation B (Figure \ref{fig:sim-b}), all estimators except $\hat\tau_\text{Saturated}$ are biased when the SEM is misspecified with $\hat\tau_\text{SEM}$ having the largest bias. However, $\hat\tau_\text{SEM}$, $\hat\tau_\text{BIC}$, and $\hat\tau_\text{SL}$ are more efficient than $\hat\tau_\text{Saturated}$ and reduce the RMSE when endpoint correlations are within 0.2 of what is assumed under a correctly specified SEM (i.e., with $s=1$ under the alternative and $s=0$ under the null).
Although $\hat\tau_\text{SEM}$ and $\hat\tau_\text{BIC}$ gain more efficiency than the $\hat\tau_\text{SL}$ in the presence of a correctly specified SEM, they have larger biases when the SEM is misspecified.
This results in inadequate coverage and inflates Type~I error rates across a range of correlations for both $\hat\tau_\text{SEM}$ and $\hat\tau_\text{BIC}$. 
However, under the global null (Figure \ref{fig:sim-b2}), all estimators are unbiased even when the SEM is misspecified. 
Moreover, $\hat\tau_\text{SEM}$, $\hat\tau_\text{BIC}$, and $\hat\tau_\text{SL}$ gain efficiency over $\hat\tau_\text{Saturated}$ regardless of model specification. 
Of note, additional precision is gained under model misspecification when $s < 1$ as $\Cov(Y_{i,1},Y_{i,2}|A_i)$ approaches zero.

\subsection{Simulation C: Binary Primary Endpoints, SEM Misspecified}

Finally, we consider a scenario with the same data generating model for the two secondary endpoints as Simulation B but with a binary primary endpoint with $\Pr(Y_{i,1}=1|A_i=0)=0.15$ and $\Pr(Y_{i,1}=1|A_i=1)=0.25$. 
Again, there is 50\% power to detect an effect on the primary endpoint with $n=250$.
Endpoints are simulated under a multivariate Gaussian model with $Y_{i,1}$ representing a dichotomized version of the latent Gaussian $Y_{i,1}^*$.
We specify 
$$\Cov(Y_{i,1}^*, Y_{i,2}, Y_{i,3}|A_i) = \begin{bmatrix} 1 & 0.51s & 0.43s \\   & 1 & 0.42 \\  & & 1 \end{bmatrix}$$ 
and vary $s \in \{0, 0.25, \ldots, 1.25\}$ to evaluate estimator performance over a range of correlations. 
When $s = 1$, the model corresponds to an SEM with $\gamma = 0.5$ and $\vec\lambda=(0.72, 0.7, 0.6)^T$; otherwise, the SEM is misspecified.

We additionally assess Type~I error control over the same correlation structures under the null hypothesis for the primary endpoint alone with respective average treatment effects of $0, 0.35$, and $0.3$ with $\Pr(Y_{i,1}|A_i)=0.15$. 
Under this null, the SEM is correctly specified with $\gamma = 0.5$ and $\vec\lambda=(0, 0.7, 0.6)^T$ if and only if $s = 0$.

\begin{figure}[htbp]
    \centering
    \includegraphics[width=\linewidth]{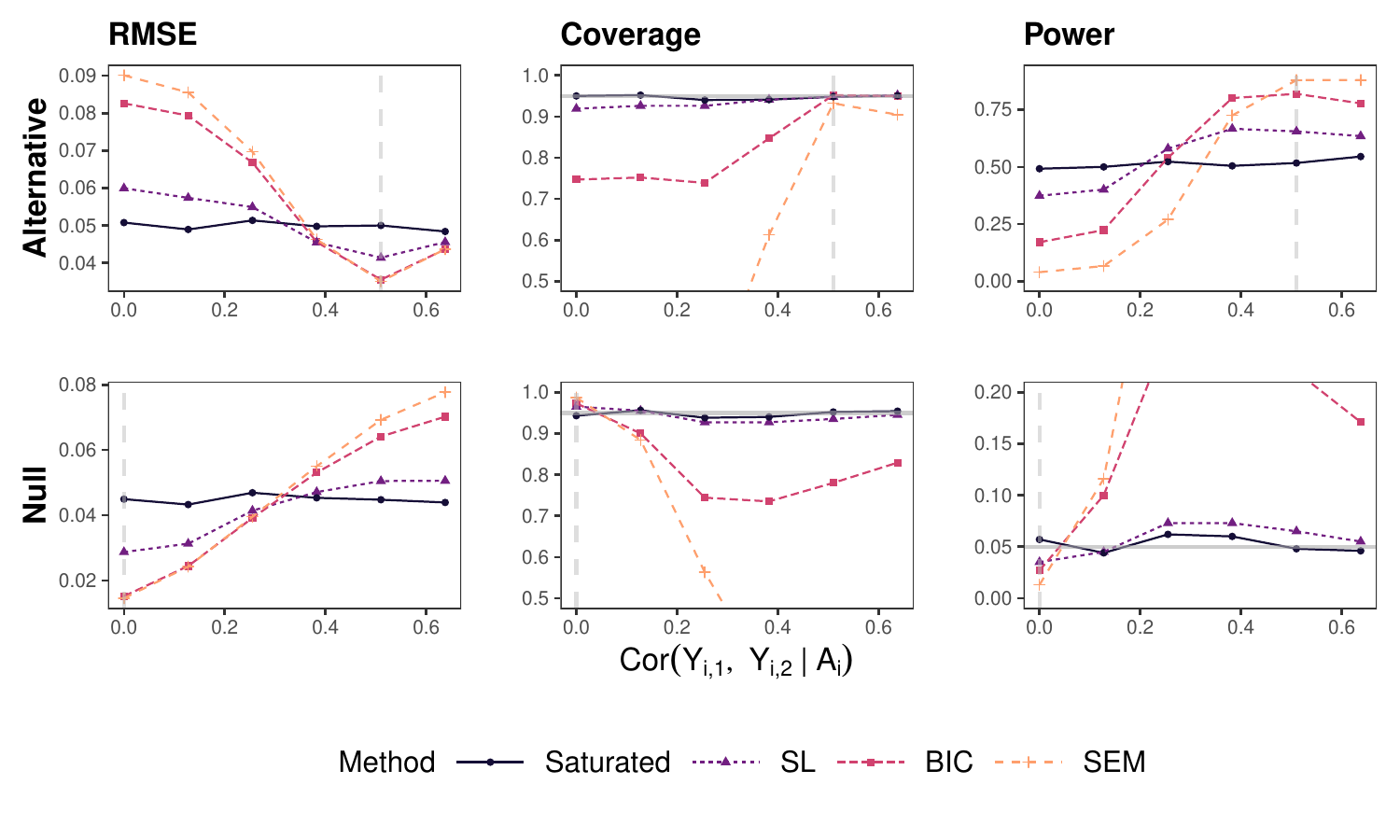}
    \caption{
        Simulation C: Root mean squared error (RMSE), coverage, and power when the structural equation model is misspecified at all but one endpoint correlation with a binary primary endpoint. 
        A dashed vertical line indicates the correlation at which the structural equation model is correctly specified.
    }
    \label{fig:sim-c}
\end{figure}

Results (Figure \ref{fig:sim-c}) are similar to what was observed with a Gaussian primary endpoint in Simulation B.
All estimators except $\hat\tau_\text{Saturated}$ are biased whenever the SEM is misspecified; however these estimators gain efficiency over $\hat\tau_\text{Saturated}$ for a range of $s$ around $s = 1$ or $s=0$ under the alternative and null hypotheses, respectively. $\hat\tau_\text{SL}$ has the highest coverage and the least Type~I error inflation among the model-averaging and SEM estimators. 

\section{Application to Tobacco Regulatory Science Research}\label{sec:application}

\citet{hatsukami_reduced_2024} recently conducted a randomized trial to evaluate VLNC cigarettes in the presence of alternative nicotine products through a virtual marketplace. 
Participants were randomly assigned into a virtual marketplace with either VLNC or NNC cigarettes in which they could exchange points for alternative nicotine products such as e-cigarettes. 
The primary endpoint was the average cigarettes smoked per day 12 weeks after randomization. 
The study randomized $n=438$ participants and found that participants randomized to the VLNC marketplace had significantly lower cigarettes smoked per day at the end of the intervention versus those randomized to the NNC marketplace.

Our analysis seeks to better understand how VLNC cigarettes affect carbon monoxide-verified ($\le$6 ppm) abstinence from smoking among Black PWS.
Because the trial only randomized $n=99$ Black PWS and because abstinence is a relatively rare binary endpoint, we do not expect that there is sufficient information to precisely estimate this treatment effect using conventional methods.
However, the study also collected data on additional biomarkers of nicotine and toxicant exposure  such as total nicotine equivalents (TNE) and cyanoethyl mercapturic acid (CEMA).
TNE (nmol/mg creatinine) is a biomarker of nicotine exposure whereas CEMA (pmol/mg creatinine) measures toxicant exposure. 
We hypothesize that these endpoints will be negatively correlated with abstinence from smoking and can be used to obtain a more precise treatment effect estimate.

We estimate the effect of VLNC cigarettes on abstinence among Black PWS using the difference in sample proportions ($\hat\tau_\text{Saturated}$) as well as by incorporating information from week 12 log-transformed TNE and CEMA using the estimators introduced in Section \ref{sec:jm}.
Missing data are handled via multiple imputations by chained equations; five independent datasets are generated and analyzed. 
Estimates for $\hat\tau_\text{Saturated}$ are pooled using Rubin's rules assuming approximate normality. 
CIs for $\hat\tau_\text{SEM}$, $\hat\tau_\text{BIC}$, and $\hat\tau_\text{SL}$ are constructed by taking the 2.5\% and 97.5\% percentiles of the pooled sample of all bootstrapped estimates across all imputations with \num[group-separator={,}]{20000} bootstrapped iterations performed for each imputed dataset \citep{schomaker_bootstrap_2018}.
We note the choice to use percentile-based inference for $\hat\tau_\text{SEM}$ as additional simulation studies suggest that the sampling distribution of $\hat\tau_\text{SEM}$ can be notably skewed with a rare binary primary endpoint and a small sample.

\begin{figure}[htbp]
    \centering
    \includegraphics[width=\linewidth]{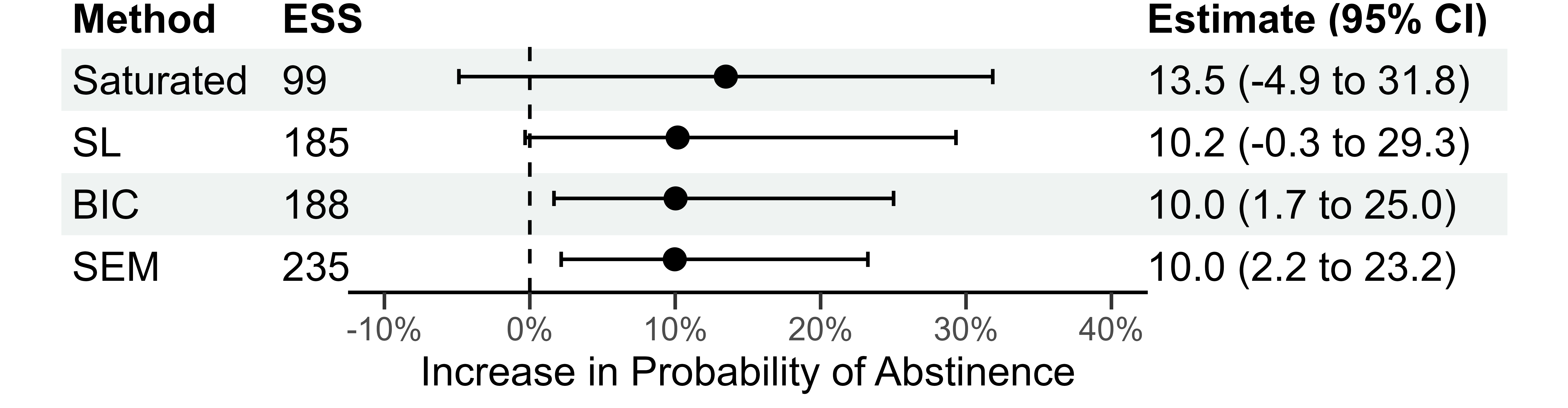}
    \caption{
        Estimated treatment effects, 95\% confidence intervals, and effective sample sizes (ESSs) for the effect of very low nicotine content cigarettes on abstinence from smoking among Black people who smoke using the considered estimators. 
    }
    \label{fig:cenic}
\end{figure}

Estimated treatment effects, 95\% CIs and effective sample sizes (ESSs) are presented in Figure \ref{fig:cenic}, where the ESS of any estimate is the ratio of its precision (inverse variance) to the saturated estimate's precision, multiplied by the sample size.
The saturated estimator estimates an increase of 13 percentage points in the probability of abstinence whereas all estimators that used information from secondary endpoints estimates an increase of 10 percentage points.
All estimates that used secondary endpoints are more precise with estimated variances approximately half that of the saturated estimator leading to ESSs approximately twice as large and smaller CIs. 
Averaged across imputations, respective weights of $\hat\omega_\text{BIC}=0.983$ and $\hat\omega_\text{SL}=0.939$ are placed on the SEM estimator.

\section{Discussion}\label{sec:discussion}

We proposed a treatment effect estimator based on a joint model that incorporates secondary efficacy endpoints to gain efficiency. 
This estimator gains precision over the standard saturated treatment effect estimator when modeling assumptions are correct but can induce substantial biases when assumptions are violated.
To mitigate this bias, we proposed an estimator that takes a weighted average of our estimator and the standard unbiased estimator.
We facilitated model averaging using both the BIC and Super Learning and found that both approaches reduced bias versus our initial estimator with Super Learning tending to lead to smaller biases while still gaining efficiency across a wide range of data generative models.

Our work is situated in the broader literature of statistical approaches for improving the efficiency of RCTs including but not limited to stratification \citep{taves_minimization_1974, pocock_sequential_1975}, covariate adjustment \citep{senn_covariate_1989, frison_repeated_1992, pocock_subgroup_2002}, and borrowing data from external sources \citep{ibrahim_power_2000, hobbs_hierarchical_2011, kaizer_bayesian_2018}. 
These approaches are often used in tandem with one another; our work could also be used in conjunction with these methods to gain additional efficiency.
Others have considered using secondary endpoints in the context of RCTs.
Two popular approaches are the use of composite endpoints \citep{freemantle_composite_2003} and the win ratio for hierarchal composite endpoints \citep{pocock_win_2012,  pocock_win_2023}.
However, both of these approaches change the estimand away from the primary outcome in undesirable ways: when using composite endpoints, the treatment effect is now a function of the risk of the primary or a secondary outcome and the win ratio reformulates the treatment effect with respect to the probability of a more desirable outcome under a specified hierarchy \citep{ferreira-gonzalez_problems_2007, tomlinson_composite_2010, bakal_applying_2015}.
This reformulation may not be in line with recent United States Food and Drug Administration guidelines on estimands \citep{us_fda_center_for_drug_evaluation_and_research_e9r1_2021}.
We recently proposed an approach to dynamic data borrowing that uses secondary endpoints \citep{wolf_leveraging_2024}.
Although this work improves estimation accuracy, it requires external data which is often not available and precision is only gained when data sources are exchangeable.
In contrast, our proposed estimator targets common estimands defined only with respect to the primary endpoint while leveraging information from secondary endpoints to gain efficiency without requiring external commensurate data.

Because we fit a model for the distribution of the primary endpoint, our method can target any functional of this distribution. 
While we have focused on the average treatment effect, one could target other common estimands related to the first moment such as a risk ratio or odds ratio, noting that targeting any higher-order moments or more complex functionals will induce more sensitivity to distributional assumptions.
Although our model can support ordinal endpoints, we note the challenges in identifying an appropriate estimand for a primary ordinal endpoint.
Several previous trials have analyzed an odds ratio assuming a proportional odds model in this setting \citep{activ-3tico_ly-cov555_study_group_neutralizing_2021, polizzotto_hyperimmune_2022-1}; we note that this estimand is only identifiable under a parametric model.
Our estimator cannot target this odds ratio because we require a different parametric model based on Gaussian latent variables; however it can target a probit regression coefficient which requires similar parametric identification.

The consistency and efficiency gains of our initial estimator are only guaranteed when the one-factor SEM is correctly specified.
This requires strong assumptions about the mean-variance relationship of all endpoints which may be violated in practice.
Additionally, we found that our proposed estimators were unbiased under the global null hypothesis.
However, we note that in simulation studies our SL estimator had acceptable performance with RMSE reduction across a wide range of correlations that violated the SEM assumptions.
The RMSE of the SL estimator only tended to be higher than the saturated estimator when the SEM was a poor fit for the data; we believe that many of these scenarios are unlikely to occur in practice under reasonable endpoint selection.
One such setting is when the treatment is efficacious on all endpoints with the secondary endpoints only weakly associated with the primary endpoint but strongly associated with each other.
The other occurred when there was a null effect on the primary endpoint but there was an efficacious effect on the secondary endpoints and all endpoints were strongly correlated. 
We note that model averaging, in particular Super Learning, further protected against this lack of fit.
For this reason, we suggest Super Learning unless there is strong \textit{a priori} evidence that the one-factor SEM would be correctly specified.

In practice, an investigator needs to decide which secondary endpoints to incorporate in their model to gain efficiency.
Ideal endpoints should both exhibit the mean-variance relationship from Equation \ref{eqn:sem-observed} to avoid inducing bias and maximize the statistical information for the treatment effect on the primary endpoint.
To identify endpoints that are reasonably modeled via Equation \ref{eqn:sem-observed}, one could consider endpoints that are hypothesized to follow the causal latent variable model found in Equation \ref{eqn:sem} or test assumptions using data from previous studies for several sets of secondary endpoints.
We found that continuous, versus binary or ordinal, secondary endpoints tend to offer the most statistical information and therefore should be the first endpoints considered.
Alternatively. for a more data-adaptive approach, one could posit multiple SEMs which include different secondary endpoints and use model averaging via Super Learning across the saturated model and all considered SEMs. 

While we facilitated model averaging using the BIC and Super Learning, additional research could explore using the focused information criterion, which assesses the performance of each model with respect to estimating a target quantity \citep{hjort_frequentist_2003}, and could be used to weigh estimators according to their mean squared error when estimating the average treatment effect. 
Alternatively, one could consider a Bayesian approach and employing shrinkage priors on the endpoint covariance matrix to facilitate a similar structure as explored by \cite{muthen_bayesian_2012}.

RCTs have the potential to generate high quality scientific evidence; however researchers are often forced to compromise between scientifically relevant yet imprecise endpoints and powerful but less interpretable endpoints. 
Our proposed estimators lead to improvements in efficiency and power versus standard treatment effect estimators.
This work adds a novel contribution to a wide literature of statistical methodology to improve the efficiency of RCTs, allowing researchers to generate high quality evidence with strong internal validity in settings where practical constraints may limit their options.

\section*{Acknowledgments}

The authors thank their collaborator, Dr. Dorothy K. Hatsukami, for providing access to the data used to illustrate their method.\vspace*{-8pt}

\section*{Funding}
This study was funded by the National Institute on Drug Abuse (Award Numbers R01DA046320 and U54DA031659) and National Center for Advancing Translational Science (Award Number UM1TR004405). The content is solely the responsibility of the authors and does not necessarily represent the official views of the National Institutes of Health and the Food and Drug Administration Center for Tobacco Products.
JMW was also supported by a University of Minnesota Data Science Initiative Graduate Assistantship with funding made available by the MnDRIVE initiative.\vspace*{-8pt}

\section*{Conflict of Interest}
None declared.\vspace*{-8pt}

\bibliography{references}  

\setcounter{equation}{0}
\setcounter{section}{0}
\setcounter{figure}{0}
\makeatletter
\renewcommand\thesection{\Alph{section}}
\numberwithin{equation}{section}
\numberwithin{figure}{section}
\makeatother

\section{Likelihood Maximization under SEM Constraints}\label{sec:sem-likelihood}

\subsection{Multivariate Gaussian Case}

When $\vec{Y}_i|\eta_i\sim N\{\vec\nu + \vec\lambda \eta_i, \operatorname{diag}(\vec\theta)\}$ and $\eta_i|A_i \sim N(\gamma A_i, 1)$, we can integrate $\eta_i$ out of the likelihood to obtain the observed data likelihood:
\begin{equation}
    \vec{Y}_i|A_i \sim N\left\{\vec\nu + \vec\lambda\gamma A_i,\operatorname{diag}(\vec\theta) + \vec\lambda\vec\lambda^T   \right\}
\end{equation}
The log likelihood is given by
\begin{align}
\begin{split}
    \ell(\vec\psi) &= -\frac{nP}{2}\log(2\pi) - \frac{n}{2}\log\left\{\left|\operatorname{diag}(\vec\theta) + \vec\lambda\vec\lambda^T\right|\right\} + {} \\
    &\quad\, -\frac12 \sum_{i=1}^n \left\{\vec{Y}_i - (\vec\nu + \vec\lambda\gamma A_i) \right\}^T \left\{\operatorname{diag}(\vec\theta) + \vec\lambda\vec\lambda^T \right\}^{-1} \left\{\vec{Y}_i - (\vec\nu + \vec\lambda\gamma A_i) \right\}. 
\end{split}
\end{align}
No closed form maximum likelihood estimator exists due to the mean-variance relationship; instead, we maximize the log likelihood using numeric methods.

\subsection{Cases With Non Gaussian Endpoints}

First, we suppose where one endpoint is binary or ordinal, and all others are Gaussian. Without loss of generality, we suppose that $Y_{i,1}$ is the binary/ordinal endpoint.
The observed data likelihood is then $\sum_{i=1}^n f_i(\vec{Y}_i|A_i;\vec\psi)$ where 
\begin{equation}
    f_i(\vec{Y}_i|A_i;\vec\psi) = f_i(Y_{i,1}|A_i,Y_{i,2},\ldots,Y_{i,P};\vec\psi) f_i(Y_{i,2},\ldots,Y_{i,P}|A_i;\vec\psi).
\end{equation}
Here, $f_i(Y_{i,2},\ldots,Y_{i,P}|A_i;\vec\psi)$ is multivariate Gaussian.
We then consider the conditional density of $Y_{i,1}^*$:
\begin{equation}
    Y_{i,1}^*|A_i,Y_{i,2},\ldots,Y_{i,P} \sim N(\mu_i, \sigma^2)
\end{equation}
where
\begin{equation}
    \mu_i = \nu_1 + \lambda_1\gamma A_i + \vec{V}_{(1, -1)} \vec{V}_{(-1, -1)}^{-1} \left\{\vec{Y}_{i(-1)} - (\vec\nu_{(-1)} + \vec\lambda_{(-1)}\gamma A_i) \right\}
\end{equation}
and
\begin{equation}
    \sigma^2 =  \vec{V}_{(1,1)} - \vec{V}_{(1, -1)} \vec{V}_{(-1,-1)}^{g} \vec{V}_{(-1, 1)}  
\end{equation}
for $\vec{V} = \operatorname{diag}(\vec\theta) + \vec\lambda\vec\lambda^T$ and where $\vec{V}_{(-1,-1)}^{g}$ is a generalized inverse of $\vec{V}_{(-1,-1)}$. 
Then, the conditional density of $Y_{i,1}$ can be given by integrating this density over the appropriate threshold values based on the value of $Y_{i,1}$. 
For example, for a binary endpoint:
\begin{align*}
    f_i(Y_{i,1}|A_i,Y_{i,1}^*;\vec\psi) &= \begin{cases} \Phi(-\mu_i/\sigma) & Y_{i,1}^* = 0 \\ 1 - \Phi(-\mu_i/\sigma) & Y_{i,1}^* = 1 \end{cases}.
\end{align*}

This can be generalized for cases with two or more non-Gaussian endpoints. Now, we integrate over the (multivariate Gaussian) joint density of $Y_{i,1}^*, \ldots, Y_{i,p}^*|Y_{i,p_1},\ldots,Y_{i,P}$ within the appropriate thresholds based on the values of $Y_{i,1},\ldots,Y_{i,p}$. 
Again, likelihood maximization is accomplished through numerical approaches.

\section{Super Learning}\label{sec:super}
We use Super Learning to obtain a weighted average between estimates of the conditional mean under the saturated and SEM models.
This section provides a summary of \cite{van_der_laan_super_2007} in the context of our estimators using notation introduced in the seminal manuscript.

Super Learning directly estimates the conditional expectation of the primary endpoint. This is accomplished by minimizing a loss function over a parameter space $\Psi$ of considered models for the conditional expectation.
Here, the true parameter is given by 
\begin{equation}
    \psi_0(a) = \E(Y_{i,1}|A_i=a)
\end{equation}
which can be seen to minimize the squared-error loss function 
\begin{equation}
    \loss(\vec{O}_i, \psi) = \{Y_{i,1}-\psi(A_i)\}^2
\end{equation}
where $\vec{O}_i=(A_i, \vec{Y}_i^T)^T$ consists of all data for subject $i$; that is, 
\begin{equation}
    \psi_0=\arg\min_{\psi\in\Psi} \E \loss(\vec{O}_i, \psi).
\end{equation}
However, other loss functions and parameters could be considered; in particular for an ordinal primary endpoint the parameter would typically be the vector of conditional probabilities $\Pr(Y_{i,1}=k|A_i=a)$ for $k=1,\ldots,K$
and the loss function would typically be the marginal log likelihood for the primary endpoint.

We will use $V$-fold cross validation to estimate the risk of any given estimator for the conditional mean, $\hat\psi_n$. 
We proceed with a brief overview of $V$-fold cross validation.
Let $\nu\in\{1,\ldots,V\}$ index data splits into testing ($T(\nu)$) and validation ($V(\nu)$) sets and let $B_n^\nu(i)=I\{i\in V(\nu)\}$ be a split indicator for fold $\nu$.
We consider a generic estimate of the parameter as a function from the sample to the parameter space: $\hat\Psi: \mathcal{M}_n \to \Psi$ where $\hat\psi_n=\hat\Psi(\P_n):\mathcal{A}\to\mathcal{Y}$ and $\hat\psi_{n,\nu}=\hat\Psi(\P_{n, T(\nu)}):\mathcal{A}\to\mathcal{Y}$ are estimates respectively fit using the empirical distributions of  the entire sample and on the training set $T(\nu)$.
The risk of an estimate is given by
\begin{equation}
    \risk(\hat\psi_n, \P) = \int \loss\{\vec{O}, \hat\Psi(\P_n)\}\dd\P.
\end{equation}
We estimate this quantity using the $V$-fold risk:
\begin{equation}
    \E_{B_n}\risk(\hat\psi_{n,\nu}, \P_{n,V(\nu)}) = \E_{B_n} \int \loss\{\vec{O}, \hat\Psi(\P_{n, T(\nu)})\}\dd\P_{n, V(\nu)}
\end{equation}
where $\P$ is the true data generative distribution function for $\vec{O}_i$ and $\P_{n, T(\nu)}$ and $\P_{n, V(\nu)}$ are the respective empirical distribution functions in the training and validation samples

Super Learning then considers a collection of candidate models to minimize this risk; in our context we have two models $\hat\psi_{n,1}$ and $\hat\psi_{n,2}$, which are estimators for the conditional mean respectively using the SEM and saturated model. 
Letting $\vec{Z}_i = (\hat\psi_{n,1,\nu(i)}(A_i), \hat\psi_{n,2,\nu(i)}(A_i))$ be a vector of the predicted values for subject $i$ using the model trained on $T\{\nu(i)\}$, Super Learning seeks to estimate $\E(Y_{i, 1}|\vec{Z}_i)=m(\vec{z}|\vec\omega)$ through a mapping $\tilde\Psi(\P_{n,\vec{Y}_1,\vec{Z}}): \mathcal{Y}^2 \to \mathcal{Y}$.
In particular, we consider meta-learners of the form $\{m(\vec{z}|\vec\omega):m(\vec{z}|\vec\omega)= \omega Z_{i,1} + (1-\omega) Z_{i, 2}, 0 \le \omega \le 1 \}$ which are indexed by $\omega\in(0,1)$.
Then letting $K(n)=|\mathcal{W}_n|$ be the number of considered grid points where $K(n)\le n^q$ for some $q<\infty$, we estimate
\begin{equation}
    \hat\omega_n = \arg\min_{\omega\in\mathcal{W}_n} \sum_{i=1}^n \{Y_{i,1} - m(\vec{Z}_i|\omega)\}^2
\end{equation}
to obtain the estimator
\begin{equation}
    \hat\psi_n(a) = m[\{\hat\psi_{n,1}(a), \hat\psi_{n,2}(a) \}|\hat\omega_n].
\end{equation}
This in turn provides the model-averaged estimates for $\E(Y_{i,1}|A_i=1)$ and $\E(Y_{i,1}|A_i=0)$ of which we take the contrast to obtain $\hat\tau_\text{SL}$.

\section{Proof of Theorem 2.1}\label{sec:pf-sem-efficiency}
We will prove efficiency by framing $\hat\tau_\text{SEM}$ and $\hat\tau_\text{Saturated}$ as functions of solutions to respective sets of quadratic and linear estimating equations.
This framework allows us to readily derive the asymptotic properties of both treatment effect estimators using M-estimation theory.

We begin by introducing additional notation: let $\vec{f}(a, \vec\mu)=\E(\vec{Y}_i|A_i=a) = \vec\alpha+\vec\beta a$ and  $\mat{V}(a, \vec\mu, \vec\xi) = \Var(\vec{Y}_i|A_i=a) =  \operatorname{diag}(\theta_1,\ldots,\theta_P)+\theta_0\vec\beta\vec\beta^T$ be models for the conditional mean and covariance where $\vec\mu=(\vec\beta^T, \vec\alpha^T)^T$ and $\vec\xi=(\theta_0, \theta_1,\ldots,\theta_P)$.
We note that this is a slightly different parameterization for the SEM than presented in the manuscript but that the two are equivalent.
That is, $\theta_0 = \gamma^{-2}$ and  $\vec\beta = \gamma\vec\lambda$.
Importantly, $\tau_1 = \mu_1$ identifies the average treatment effect.
Then, we consider the following quadratic estimating equations for $(\vec\mu^T, \vec\xi^T)$:

\begin{equation}
    \label{eq:sem-est-sum}
    \sum_{i=1}^n 
    \begin{bmatrix}
        \mat{X}_i^T(\vec\mu) & \mat{B}_i^T(\vec\mu, \vec\xi) \\ \mat0 & \mat{E}_i^T(\vec\mu,\vec\xi)
    \end{bmatrix}
    \begin{bmatrix}
        \mat{V}_i(\vec\mu, \vec\xi, A_i) & \mat0 \\ \mat0 & \mat{Z}_i(\vec\mu, \vec\xi)
    \end{bmatrix}^{-1}
    \begin{bmatrix}
        \vec{Y}_i - \vec{f}(A_i, \vec\mu) \\ \vec{u}_i - \vec\upsilon_i(\vec\mu,\vec\xi)
    \end{bmatrix}
    =\vec0 
\end{equation}
where
\begin{align}
    \vec{u}_i &= \vech[\{\vec{Y}_i - \vec{f}(A_i, \vec\mu)\}^T\{\vec{Y}_i - \vec{f}(A_i, \vec\mu)\}], \\
    \vec\upsilon_i(\vec\mu,\vec\xi) &= \vech\{\mat{V}_i(\vec\mu, \vec\xi, A_i)\}, \\
    \mat{Z}_i(\vec\mu,\vec\xi) &= \Var(\vec{u}_i-\vec\upsilon_i), \\
    \mat{X}_i(\vec\mu) &= \pdv{\vec{f}_i}{\vec\mu}, \\
    \mat{B}_i(\vec\mu,\vec\xi) &= \pdv{\vec\upsilon_i}{\vec\mu}, \\
    \intertext{and}
    \mat{E}_i(\vec\mu,\vec\xi) &= \pdv{\vec\upsilon_i}{\vec\xi}.
\end{align}
We note that, in general, the elements of $\mat{Z}_i(\vec\mu,\vec\xi)$ are obtained by specifying of the  the third and fourth moments of $\vec{Y}_i|A_i$; here we assumed multivariate normality which allows for derivation of these moments from the mean and covariance. 
Then, \ref{eq:sem-est-sum} can be rewritten as
\begin{equation}
    \label{eq:sem-est}
    \begin{bmatrix}
        \mat{X}^T & \mat{B}^T \\ \mat0 & \mat{E}^T
    \end{bmatrix}
    \begin{bmatrix}
        \mat{V} & \mat0 \\ \mat0 & \mat{Z}
    \end{bmatrix}^{-1}
    \begin{bmatrix}
        \vec{Y} - \vec{f} \\ \vec{u} - \vec\upsilon
    \end{bmatrix}
    =\vec0 
\end{equation}
for
\begin{align}
    \vec{Y}^T &= \begin{bmatrix}
        \vec{Y}_1^T & \cdots & \vec{Y}_n^T
    \end{bmatrix}, \\ 
    \vec{f}^T &= \begin{bmatrix}
        \vec{f}_1(A_i,\vec\mu)^T & \cdots & \vec{f}_n(A_i,\vec\mu)^T
    \end{bmatrix}, \\
    \vec{u}^T &= \begin{bmatrix}
        \vec{u}_1^T & \cdots & \vec{u}_n^T
    \end{bmatrix}, \\
    \vec{\upsilon}^T &= \begin{bmatrix}
        \vec{\upsilon}_1(\vec\mu, \vec\xi)^T & \cdots & \vec{\upsilon}_n(\vec\mu, \vec\xi)^T
    \end{bmatrix}, \\
    \mat{V} &= \operatorname{blockdiag}\{\mat{V}_1(\vec\mu, \vec\xi, A_n), \ldots, \mat{V}_n(\vec\mu, \vec\xi, A_n)\}, \\
    \mat{Z} &= \operatorname{blockdiag}\{\mat{Z}_1(\vec\mu, \vec\xi), \ldots, \mat{Z}_n(\vec\mu, \vec\xi)\}, \\
    \mat{X}^T &= \begin{bmatrix}
        \mat{X}^T_1(\vec\mu) & \cdots & \mat{X}^T_n(\vec\mu)
    \end{bmatrix}, \\
    \mat{B}^T &= \begin{bmatrix}
        \mat{B}^T_1(\vec\mu,\vec\xi) & \cdots & \mat{B}^T_n(\vec\mu,\vec\xi)
    \end{bmatrix}, \\
    \intertext{and}
    \mat{E}^T &= \begin{bmatrix}
        \mat{E}^T_1(\vec\mu,\vec\xi) & \cdots & \mat{E}^T_n(\vec\mu,\vec\xi)
    \end{bmatrix},
\end{align}
where dependence of these functions on $\vec\mu$ and $\vec\xi$ has been omitted for visual clarity.
It follows that 
$(\hat{\vec\mu}_\text{Quad}^T, \hat{\vec\xi}_\text{Quad}^T)^T$ is asymptotically normal with mean $({\vec\mu}_\text{Quad}^T, {\vec\xi}_\text{Quad}^T)^T$ and variance $\mat\Sigma_\text{Quad}$
where
\begin{align}
    \mat\Sigma_\text{Quad}^{-1} &= \Var\left\{\begin{bmatrix}
        \mat{X}^T & \mat{B}^T \\ \mat0 & \mat{E}^T
    \end{bmatrix}
    \begin{bmatrix}
        \mat{V} & \mat0 \\ \mat0 & \mat{Z}
    \end{bmatrix}^{-1}
    \begin{bmatrix}
        \vec{Y} - \vec{f} \\ \vec{u} - \vec\upsilon
    \end{bmatrix} \right\} \\
    \begin{split}
    &= \begin{bmatrix}
        \mat{X}^T & \mat{B}^T \\ \mat0 & \mat{E}^T
    \end{bmatrix}
    \begin{bmatrix}
        \mat{V} & \mat0 \\ \mat0 & \mat{Z}
    \end{bmatrix}^{-1}
    \Var\left\{\begin{bmatrix}
        \vec{Y} - \vec{f} \\ \vec{u} - \vec\upsilon
    \end{bmatrix}\right\} 
    \begin{bmatrix}
        \mat{V} & \mat0 \\ \mat0 & \mat{Z}
    \end{bmatrix}^{-1}
    \begin{bmatrix}
        \mat{X} & \mat0 \\
        \mat{B} & \mat{E}
    \end{bmatrix}
    \end{split}\\
    &= \begin{bmatrix}
        \mat{X}^T & \mat{B}^T \\ \mat0 & \mat{E}^T
    \end{bmatrix}
    \begin{bmatrix}
        \mat{V} & \mat0 \\ \mat0 & \mat{Z}
    \end{bmatrix}^{-1}
    \begin{bmatrix}
        \mat{X} & \mat0 \\
        \mat{B} & \mat{E}
    \end{bmatrix} \\
    \label{eq:sigma-to-invert}
    &= \begin{bmatrix}
        \mat{X}^T\mat{V}^{-1}\mat{X} + \mat{B}^T\mat{Z}^{-1}\mat{B} & \mat{B}^T\mat{Z}^{-1}\mat{E} \\
        \mat{E}^T\mat{Z}^{-1}\mat{B} &
        \mat{E}^T\mat{Z}^{-1}\mat{E}
    \end{bmatrix}.
\end{align}
(We note that we have assumed a fixed design so that $\mat{X}$, $\mat{B}$, and $\mat{E}$ can be treated as fixed quantities for simplicity; the proof may be generalized to random designs through additional notation.)
The asymptotic variance of $\hat{\vec\mu}$ is given by the upper $|\vec\mu|\times|\vec\mu|$ of $\mat\Sigma_\text{Quad}$, $\mat\Sigma_{\text{Quad}, \vec\mu}$ 
Inverting \ref{eq:sigma-to-invert}, we find that the desired matrix is
\begin{align}
     \mat\Sigma_{\text{Quad}, \vec\mu} &= \left\{\mat{X}^T\mat{V}^{-1}\mat{X} + \mat{B}^T\mat{Z}^{-1}\mat{B}-\mat{B}^T\mat{Z}^{-1}\mat{E}(\mat{E}^T\mat{Z}^{-1}\mat{E})^{-1}\mat{E}^T\mat{Z}^{-1}\mat{B} \right\}^{-1} \\
     &= \left[\mat{X}^T\mat{V}^{-1}\mat{X} + \mat{B}^T\mat{Z}^{-1/2}\left\{\mat{I} -   \mat{Z}^{-1/2}\mat{E}(\mat{E}^T\mat{Z}^{-1}\mat{E})^{-1}\mat{E}^T\mat{Z}^{-1/2}\right\}\mat{Z}^{-1/2}\mat{B} \right]^{-1} \\
     \intertext{Letting $\mat{G} = \mat{Z}^{-1/2}\mat{E}$}
     &= \left[\mat{X}^T\mat{V}^{-1}\mat{X} + \underbrace{\mat{B}^T\mat{Z}^{-1/2}\left\{\mat{I} - \mat{G}(\mat{G}^T\mat{G})^{-1}\mat{G}^T  \right\}\mat{Z}^{-1/2}\mat{B}}_{= \mat{A}} \right]^{-1}.
\end{align}
It can be seen that $\mat{A}$ is positive semi-definite because $\mat{I} - \mat{G}(\mat{G}^T\mat{G})^{-1}\mat{G}^T$ is a projection matrix. It follows that
\begin{equation}
    \left[\mat{X}^T\mat{V}^{-1}\mat{X} + \mat{A} \right]^{-1} \le (\mat{X}^T\mat{V}^{-1}\mat{X})^{-1}
\end{equation}
Here, 
${\mat\Sigma}_{\text{Lin},\vec\mu}=(\mat{X}^T\mat{V}^{-1}\mat{X})^{-1}$
is the asymptotic variance of $\hat{\vec\mu}_\text{Lin}$ under the linear estimating equations (Equation \ref{eq:sem-est} with $\mat{B}=\mat0$) that independently estimate the conditional mean and variance.
It follows that
\begin{align}
    {\mat\Sigma}_{\text{Quad},\vec\mu} &\le {\mat\Sigma}_{\text{Lin},\vec\mu}
    \intertext{and}
    \AVar(\hat\tau_\text{SEM}) &\le \AVar(\hat\tau_\text{Saturated}).
\end{align}

\section{Proof of Theorem 2.2}\label{sec:pf-ma-consistent}

First we consider $\hat\tau_\text{BIC}$ where
\begin{equation}
    \hat\omega_\text{BIC} = \left[\exp\left\{\frac12\left(\BIC_\text{SEM}-\BIC_\text{Saturated} \right) \right\} +1\right]^{-1}.
\end{equation}
It is thus sufficient to consider 
\begin{align}
    \BIC_\text{SEM}-\BIC_\text{Saturated} &= \{-2\ell_\text{SEM}+k_\text{SEM}\log(n)\} - \{-2\ell_\text{Saturated}+k_\text{Saturated}\log(n)\} \\
    &= \lambda + (k_\text{SEM}-k_\text{Saturated})\log(n)
\end{align}
where $\ell_\text{SEM}$ and $\ell_\text{Saturated}$ are the log likelihood values under the respect MLEs and $k_\text{SEM}=3P+1$ and $k_\text{Saturated}=2P+\frac12P(P+1)$ are the models' respective numbers of parameters.
We note that $\lambda$ is a likelihood ratio test statistic, which will be used to derive this difference's limiting behavior.
If the SEM is correctly specified, $\lambda \dto \chi^2(k_\text{SEM}-k_\text{Saturated}, 0)$
where $\chi^2(a,b)$ denotes a chi-squared distribution with degrees of freedom $a$ and noncentrality parameter $b$.
It follows that 
\begin{equation}
    \log(n)^{-1}\{\lambda + (k_\text{SEM}-k_\text{Saturated})\log(n)\} \pto k_\text{SEM}-k_\text{Saturated} < 0 
\end{equation}
and $\hat\omega_\text{BIC}\pto1$.
Consequentially, $\hat\tau_\text{BIC}\pto\hat\tau_\text{SEM}$ where, given that the SEM is correct $\hat\tau_\text{SEM}\pto\tau_1$.
However, if the SEM is misspecified, $\lambda$ is approximately distributed as $\chi^2\{k_\text{SEM}-k_\text{Saturated}, n \delta\}$
where 
\begin{equation}
    \delta = 2\left[\E\{\ell_i(\tilde{\vec\psi}_\text{SEM})\}-\E\{\ell_i(\tilde{\vec\psi}_\text{Saturated})\} \right]
\end{equation}
is twice the expected difference in log likelihoods for a single observation under the saturated model and SEM using the pseudo-true and true parameter values. 
Specifically, the pseudo-true value of $\tilde{\vec\psi}_\text{SEM}$ is given by
\begin{equation}
    \tilde{\vec\psi}_\text{SEM} = \arg \max_{\vec\psi\in\vec\Psi} \E\left\{\log f_i(\vec{Y}_i|A_i, \vec\psi)\right\}.
\end{equation}
It can then be seen that
\begin{equation}
    \{2(k_\text{SEM}-k_\text{Saturated})+4n\delta\}^{-1}\{\lambda + (k_\text{SEM}-k_\text{Saturated})\log(n)\} \pto  \frac14.
\end{equation}
Consequently, $\hat\omega_\text{BIC}\pto0$ and $\hat\tau_\text{BIC}\pto\hat\tau_\text{Saturated}$ where $\hat\tau_\text{Saturated}\pto\tau_1$.

Now we consider $\hat\tau_{SL}$. 
Established theory \citep{van_der_laan_super_2007} states that $\hat\psi_n(a)$ will be asymptotically equivalent to the oracle estimator where the oracle estimator is taken by using the oracle selector 
\begin{equation}
  \tilde\omega_n = \arg \min_{\omega\in\mathcal{W}_n}\frac1V \sum_{\nu=1}^V d\left\{ \hat\Psi_\omega(\P_{n,T(\nu)}), \psi_0 \right\},  
\end{equation}
where
\begin{equation}
    d(\psi, \psi_0) = \E\left\{\loss(A_i, \psi) - \loss(A_i,\psi_0) \right\} = \E\left\{\psi(A_i) - \psi_0(A_i) \right\}^2.
\end{equation}
If the SEM is correctly specified, $\AVar\{\hat\psi_1(a)\} \le \AVar\{\hat\psi_2(a)\}$ and $\tilde\omega_n\to 1$. In order for $\hat\psi_n(a)$ to have equivalent asymptotic performance, it must be the case that $\hat\omega_\text{SL}\to 1$ as well. 
As a result, Hence, $\hat\tau_\text{SL}\pto\hat\tau_\text{SEM}$.
Inversely, if the SEM is misspecified 
\begin{align}
    d\left(\hat\psi_1, \psi_0 \right) &= \E\left[\E\left\{\psi_1(A_i) - \psi_0(A_i)\right\}^2|A_i \right] \\
    &= \E\left[{\underbrace{\E\{\hat\psi_1(A_i) - \psi_0(A_i)|A_i\}}_{\text{Bias}}}^2 + \underbrace{\Var\{\hat\psi_1(A_i)|A_i\}}_{\text{Variance}} \right]
\end{align}
It can be shown using techniques introduced in the previous proof that this bias term converges to some nonzero constant, 
\begin{equation}
    \E\{\hat\psi_1(A_i) - \psi_0(A_i)|A_i\} \pto b_{A,1},
\end{equation}
whereas the variance term tends to zero at a $|T(\nu)|^{-1}$ rate,
\begin{equation}
    |T(\nu)|^{-1}\Var\{\hat\psi_1(A_i)|A_i\} \pto 0
\end{equation}
However, the saturated estimator's distance converges to zero as the model results in unbiased point estimation such that
\begin{equation}
    \E\{\hat\psi_2(A_i) - \psi_0(A_i)|A_i\} \pto b_{A,2} = 0,
\end{equation}
where the variance also goes to zero at a $|T(\nu)|^{-1}$ rate:
\begin{equation}
    |T(\nu)|^{-1}\Var\{\hat\psi_2(A_i)|A_i\} \pto 0.
\end{equation}
It follows that
\begin{equation}
    d\left(\hat\psi_1, \psi_0 \right)-d\left(\hat\psi_2, \psi_0 \right) \pto b_{1,0}^2(1-\pi) + b_{1,1}^2\pi
\end{equation}
as $|T(\nu)|\to\infty$ where $\pi = \Pr(A_i=1)$.
Consequently, the saturated model minimizes the distance function and $\tilde\omega_n\pto0$.
Again, by the asymptotic equivalence of the Super Learner and the oracle estimator, it must be the case that $\hat\omega_\text{SL}\pto0$ in order to have the same asymptotic risk. 
Hence, $\hat\tau_\text{SL}\pto\hat\tau_\text{Saturated}$.

\end{document}